\documentclass{svjour3}                     % onecolumn (standard format)

\pdfoutput=1
\smartqed  
\usepackage{graphicx}
\usepackage{times}
\usepackage{courier}
\usepackage{amsmath}
\usepackage{amsbsy}
\usepackage{amsfonts}

\begin{document}

%\title{A Fast Spatially Balanced Sample Selection Criterion}
\title{Fast Selection of Spatially Balanced Samples}
%\title{A Fast Algorithm of Spatially Balanced Sampling}
\titlerunning{Fast Selection of Spatially Balanced Samples}    
\author{F. Piersimoni \and R. Benedetti}
\institute{F. Piersimoni \at
              Istat, Directorate for Methodology and Statistical Process Design, Via Cesare Balbo 16, Rome, IT-00184, Italy \\
              \email{piersimo@istat.it}  
           \and
           R. Benedetti \at
              ``G. d'Annunzio'' University, Department of Economic Studies (DEc), Viale Pindaro 42, Pescara, IT-65127, Italy \\
              Tel.: +39-085-45083210, +39-085-45083229\\
              Fax: +39-085-45083208\\
              \email{benedett@unich.it}
}


\maketitle

\begin{abstract}
Sampling from very large spatial populations is challenging. The solutions suggested in recent literature on this subject often require that the randomly selected units are well distributed across the study region by using complex algorithms that have the feature, essential in a design--based framework, to respect the fixed first--order inclusion probabilities for every unit of the population. The size of the frame, $N$, often causes some problems to these algorithms since, being based on the distance matrix between the units of the population, have at least a computational cost of order $N^2$. In this paper we propose a draw--by--draw algorithm that randomly selects a sample of size $n$ in exactly $n$ steps, updating at each step the selection probability of not--selected units depending on their distance from the units already selected in the previous steps. The performance of this solution is compared with those of other methods derived from the {\it spatially balanced sampling} literature in terms of their root mean squared error (RMSE) using the simple random sampling (SRS) without replacement as benchmark. The fundamental interest is not only to evaluate the efficiency of a such different procedure, but also to understand if similar results can be obtained even with a notable reduction in the computational burden needed to obtain more efficient sampling designs. Repeated sample selections on real and simulated populations support this perspective. An application to the Land Use and Land Cover Survey (LUCAS) 2012 data--set in an Italian region is presented as a concrete and practical illustration of the capabilities of the proposed sample selection method.
.
\end{abstract}

\section{Introduction}
\label{s:intro}
Surveys are routinely used to gather data for environmental and ecological research. The units to be observed are often randomly selected from a population made up of geo-referenced units, and the spatial distribution of these units is information that can be used in sample design. It represents a source of auxiliaries that can be helpful to design an effective random selection strategy, which, by a proper use of this particular information and assuming that the observed phenomenon is related with the spatial features of the population, could lead to a remarkable efficiency gain of a design--based estimator of the unknown total of a {\it target} variable.\\
In recent years, with the widespread use of the Global Positioning System (GPS) technology, of remotely sensed data, of automatic address geocoding systems and of Geographical Information Systems (GIS) databases, spatial modeling and analysis of geo-referenced data-sets have received more and more attention. The use of this particular type of data is becoming popular in several research areas including geological and environmental sciences \cite{b66}, ecology \cite{b53}, cartography \cite{b40}, public health \cite{b16} and so on. Some extended reviews of current research tendencies in spatial modeling and analysis are presented in \cite{b12,b21,b26,b28,b43,b45,b52}.\\
At the same time it is surprising and incomprehensible the reason why, except for some attractive references on the subject \cite{b06,b47}, at least the same interest was not devoted to basic topics as spatial data collection and sample surveys selected from spatial populations, with the consequence that none of these authoritative texts addresses, unless mentioned as a border issue, the problem of sampling from spatial data--sets.\\
To better understand the increase in the importance of spatial data, it is also appropriate to realize that is more and more frequent the practice that National Statistical Offices geo-reference their sampling frames of physical or administrative bodies, used for social and economic surveys, not only according to the codes of a geographical nomenclature, but also adding information regarding the exact, or estimated, position of each record. With the effect that such methodological problems are no longer felt only in local and occasional experiments, although of a high scientific level, but also in economic and social surveys that provide basic statistical figures at national and regional level. Indeed area frame surveys \cite{b06} often supported, or even replaced, the traditional methods based on list frames mainly due to the absence of coverage errors and the low probability of non--responses that they guarantee. Moreover the latest fashion in official statistics, currently considered a keyword, is the statistical use of administrative data that implies the need of estimating the coverage of administrative lists, which is often carried out through area frame surveys. Spatial surveys also present some drawbacks with regard to the reduced information that can be collected in a survey involving only a direct observation and not a questionnaire to be filled. Within this context, in \cite{b02} it is claimed that area frame surveys will play a very important role in the future development of agricultural surveys in particular in developing countries.\\
In addition big spatial data--sets are very common in scientific problems, such as those involving remote sensing of the earth by satellites, climate-model output, small-area samples from national surveys, and so forth. The term {\it Big Data} includes data characterized not only by their large volume, but also by their variety and velocity, the organic way in which they are created, and the new types of processes needed to analyze them and make inference from them. The change in the nature of the new types of data, their availability, and the way in which they are collected and disseminated is fundamental. This change constitutes a paradigm shift for survey research. There is great potential in {\it Big Data}, but there are some fundamental challenges that have to be resolved before their full potential can be realized. A large data--set entails not only computational but also theoretical problems as it is often defined on a large spatial domain, so the spatial process of interest typically exhibits local homogeneity over that domain. The availability of massive spatial data that is becoming quite common emphasizes the need for developing new and computationally efficient methods tailored to handle such big data--sets. These problems have already been widely treated from a modeling approach and there is a lot of literature on the subject \cite{b13,b14,b38,b61}. On the contrary, the aspect of data collection, in particular the sample selection, has not yet been fully addressed.\\
The situation becomes even more difficult to manage if we consider that these data have complex structures. The use of points, lines, polygons or regular grids to represent spatial units and their relationships is certainly more than an established practice, it is the way used in geography to simplify reality. Each data type in this list is represented by different modes and rules representing a major framework for the inherent complexity of the analysis of spatial phenomena. This complexity is related to the stage of data collection, in that various situations can arise for recording and describing a given phenomenon under alternative topologies.\\
Reflecting on the evidence that spatial sampling algorithms can be very slow when the data--sets are made up of a large amount of data, our main purpose was to propose a fast method for random selection of units that are well spread in every dimension that can therefore be used in massive surveys without creating computational problems and that, by founding its ground on an attempt to set the concepts used by different approaches to the problem, have the flexibility deriving from a model--based (MB) approach and can be safely used even by the most strong supporters of sampling from finite populations within a design--based (DB) framework.\\
The plan of the paper is as follows. Some preliminaries and basic concepts are given in Section \ref{s:basic} together with some motivation to spread the sample over a spatial population. After a review of the literature on {\it spatially balanced samples}, in Section \ref{s:hpwd} is introduced a fast algorithm that, selecting well--spread samples, seeks to exploit the spatial characteristics of the population. In Section \ref{s:compar} the performance of the proposed method is empirically compared with some reviewed competitors by a design--based simulation study. The efficiency appraisal is evaluated in terms of the root mean squared error (RMSE) of the estimates by using the simple random sampling (SRS) as benchmark. Different estimation scenarios have been assumed for varying sample sizes and with respect to real and artificial populations characterized by different spatial distributions. Finally, in Section \ref{s:conc} the main results of the paper are discussed and some addresses for further research are provided.
\section{Basic concepts and preliminaries}
\label{s:basic}
The adoption of a spatial model exploiting the topological nature of geographical entities can be formally tackled by introducing the following {\it super--population} model for spatial data \cite{b12}: let $i \in \mathbb{R}^d$ be a generic data location in a $d$--dimensional Euclidean space within a region $U \subseteq \mathbb{R}^d$ and $Y\bigl(i\bigr)$ a random variable representing the potential datum for the {\it attribute} or {\it target} variable at spatial location $i$. Now let $i$ vary over index set $U$ so as to generate the random field $\{ Y(i):i \in U \}$. Within this model, space is featured in every respect by the set $U$, including topology and coordinates system, and location in $U$ indexed by $i$. Notice that the data generating process can lie also over a continuous spatial domain even if, to simplify the problem, it is observed only in a selection, possibly made at random, of fixed points or averaged over a selection of predefined polygons. Even if in this paper we will focus our attention on situations of this kind, which can be led back to usual finite population sampling, it is appropriate to underline that in natural resources monitoring and estimation, however, they cover an important part of all the possible sampling problems that arise in this field. There is a huge list of phenomena that can be observed in any site of a linear object, such as a river, or of a surface as it is for meteorological data. In these cases the resulting sample is a set of points or polygons whose possible positions are not predefined but chosen from an infinite set of possible sites.\\
A spatial finite population $U = \{1,\dots,i,\dots,N\}$ of size $N$ is thus recorded on a frame and from the geographical position and topology of each unit $i$ we can derive, according to some distance definition, a matrix $\mathbf{D}_U=\{d_{ij};i=1,\dots,N;j=1,\dots,N\}$ that specifies how far are all the pairs of units in the population.\\
The use of the matrix $\mathbf{D}_U$ as a synthesis of the spatial information implies the hypothesis that the dependence does not change with the position of the unit $i$ and the direction, i.e. that the random field $Y\bigl(i\bigr)$ is {\it homogeneous} and {\it isotropic} \cite{b12}, i.e. its distribution does not change if we shift or rotate the space of the coordinates. $\mathbf{D}_U$ is a very important tool to emphasize the importance to spread the sample over $U$, a property that can be related, through a variogram \cite{b12}, to the spatial dependence of $Y\bigl(i\bigr)$ but also to some form of similarity between adjacent units as a spatial clustering or a spatial stratification. If the units are points, for the definition of $\mathbf{D}_U$, we can simply resort to simple concepts of distance between sets of coordinates, but if they are polygons we should use as a distance the notion of contiguity between areal units, or it would be better to use the order of contiguity, unless we want to transform polygons into points by identifying them with their centroids. As far as linear topology is concerned, a distance measurement can not ignore assumptions about the distances between the sets of points that make up the lines. Some synthetic indexes such as the minimum and the maximum are always available while for the average, as it is necessary to calculate the area between the two lines, serious computational problems could arise.\\
A traditional objective of most surveys is estimation of the total $t_Y=\sum_{i \in U}Y\bigl(i\bigr)$ when we assume that $U$ is a finite population that becomes $t_Y=\int_U Y\bigl(i\bigr)\:d\:i$ if $U$ is not finite but is a continuous surface over $U$. In order to estimate $t_Y$, a sample of units from $U$ is selected by identifying their labels $i$ on the frame, and then measures $y_i$ of their corresponding values are collected. Considering only without replacement samples, a convenient way to state this sample selection process is to assume that, for each unit $i$ on the frame, a random vector $S=\{s_i; i=1,\dots,N\}$ is generated that is equal to 1 if a unit is selected in the sample and 0 otherwise. If the sample size is fixed then $n=\sum_{i=1}^Ns_i$.  The distribution $P\bigl(S\bigr)$, whose support is $\mathcal{S}$ and in principle is under the complete control of the sampler, defines what is generally referred to as the {\it design} of the sample survey. The first--order inclusion probability that the unit $i$ will be included in the sample is denoted by $\pi_i=\sum_{S \ni i}P(S)$, where the term $S \ni i$ means that the sum is extended over those samples that contain $i$, while the second order inclusion probability for the units $i$ and $j$ is denoted as $\pi_{ij}=\sum_{S \ni \{i,j\}}P(S)$.\\
Within the DB estimation framework we consider $P(S)$ as the only source of randomness, thus assuming that $y_i$ is not a random field and is not affected by any measurement error, we can define the well-known, and widely used in practical applications, Horvitz and Thompson (HT) \cite{b51} estimator for $t_Y$ as $\hat{t}_{HT,Y}=\sum_{i \in S}{y_i}/{\pi_i}$, where the term $i \in S$ here means that the sum is extended over those units $i$ that have been selected in the sample $S \in \mathcal{S}$. The equivalent of the HT estimator, and its properties, for continuous populations was extensively treated by \cite{b15}.\\
The alternative MB estimation framework consists in assuming that the actual finite population $y_i$ is only one of the possible realizations of the random field $Y\bigl(i\bigr)$. The population total can be decomposed as the sum of two components $t_Y=t_{Y_S}+t_{Y_{\bar{S}}}$, where $Y_S$ is the set of $y_i$ values observed in the sample while $Y_{\bar{S}}$ are the not--observed values in the remaining units of the population ($\bar{S} \equiv U-S$). Once completed the data collection $t_{Y_S}$ is known, and the estimation problem can be reduced to predict $t_{Y_{\bar{S}}}$ with $\hat{t}_{Y_{\bar{S}}}$ as the sum of the predicted values $\hat{y}_i$ arising from a suitable model $\xi$ fitted to the observed data \cite{b10,b62}.\\
It follows that, depending on which estimation approach we adopt, we should evaluate the expected value and the mean squared error (MSE) of the estimators following different criteria, when dealing with a DB approach $E_S\bigl[\bigl(\hat{t}_{HT,Y}-t_Y\bigr)\bigr]$ and $E_S\bigl[\bigl( \hat{t}_{HT,Y}-t_Y\bigr)^2\bigr]$ should be used while $E_{\xi}\bigl[\bigl( \hat{t}_{HT,Y}-t_Y\bigr)|S\bigr]$ and $E_{\xi}\bigl[\bigl( \hat{t}_{HT,Y}-t_Y\bigr)^2|S\bigr]$ are more appropriately applied when working within a MB approach, where $E_S$ and $E_{\xi}$ denote respectively expectation with regard to the sample $S \in \mathcal{S}$ and to the model $\xi$.\\
A vector of $C$ covariates $\vec{x}_i=\{x_{i1},\dots,x_{ic},\dots,x_{iC}\}$ is usually available for each unit $i$ in the frame represented by the coordinates of the unit (if they are points or centroids of polygons), the land use derived from a map, the elevation, remotely sensed data, administrative data and so on. Considering that we are dealing with spatial data the practice is that the model $\xi$ consists of two parts, a trend component that is used to relate the outcome $Y\bigl(i\bigr)$ with the auxiliaries $\vec{x}_i$ and an autocorrelation component that takes advantage of the knowledge of $\mathbf{D}_U$ to fit a variogram to the observed data. This model is known as {\it kriging} \cite{b12,b21} and, considering its theoretical properties and its popularity in practical applications, it is surely one the best candidate for $\xi$.\\
To select $S$ the logic of the optimal design is conceived so that preferential sampling \cite{b20} can be used allowing explicitly for the minimization of a criterion $\Phi$ linked with some summary statistics $Q$, usually an utility function, arising from model $\xi$ \cite{b65}. More formally the set of $n$ units that constitute the sample $S$, and possibly a set of weights $F$ associated to the sample, are the result of an optimization problem such as $\max\limits_{S,F}\bigl[Q\bigl(\xi\bigr)\bigr]$ \cite{b46,b48,b48a}. This is a combinatorial optimization problem that can be solved through some heuristics or by using the well known {\it Simulated Annealing} algorithm that has shown to provide promising results \cite{b04}. Alternatively a design criterion motivated from Bayesian learning was suggested and estimated by MCMC \cite{b25} or, if estimation of $\Phi$ is complex as the likelihood is intractable, approximate bayesian computation showed to be a feasible solution \cite{b33,b34}.\\
In a MB perspective, the concern is necessarily in finding the sample configuration that is the best representative of $U$ maximizing an objective function defined over the whole set $\mathcal{S}$ of possible samples which, in a spatial context, will surely depend on the loss of information due to spatial autocorrelation. The resulting optimal sample is selected with certainty and is of course not advised, if we assume the randomization hypothesis that is the background for DB inference \cite{b50,b60,b64}. Moreover, as clearly stated by \cite{b13}: {\it ``A model-based approach seems therefore necessary under preferential sampling, such as assuming a spatial-statistical model for $s(\cdot)$. However, this does not mean that the basic design-based notions of randomisation, stratification, and clustering cannot be used in a preferential--sampling approach, since they are all useful tools that lead to a better representation of a heterogeneous population. In particular, what happens when the model $\dots$ does not adequately describe the spatial variability in $y(\cdot)$ and $s(\cdot)$? The optimality $\dots$, and the validity of any consequent inference depends critically on the appropriateness of this model''}.\\
The answer to these questions is, of course, a highly debated topic not only for theoretical but also practical reasons, starting with the evidence that if we do not assume that a model holds for the measurement errors, the potential observations over each unit of the population, being of a deterministic nature, cannot be considered as dependent. Thus it is clear that sampling schemes to be reasonably adopted for spatial units in a DB context need a suitable alternative to the concept of spatial dependence.\\
A useful property of random samples, that of being {\it spatial balanced}, has been introduced for this purpose by \cite{b55}. To define a {\it spatial balance index} (SBI) the use of Voronoi polygons is required for any sample $S$. Such a polygon, for the sample unit $i$, includes all population units closer to $i$ than to any other sample unit $j$. If we let $\nu_i$ be the sum of the $\pi_v$ of all units $v$ in the $i$-th Voronoi polygon, for any sample unit we have $E(\nu_i)=1$. Thus the SBI:
\begin{equation}
\label{f:indexsb}
SBI(S)=Var \left ( \nu_i \right )=\frac{\sum _{i\in S}\left ( \nu_i -1 \right )^2}{n},
\end{equation}
can be used as a measure of difference from the state of perfect {\it spatial balance} (SBI(S)=0). A sample $S$ is considered {\it spatially balanced} when SBI(S) $\leq \omega_{sb}$ as every $\nu_i$ should be close to 1, where $\omega_{sb}$ is an acceptable upper bound. Note that the role played by SBI is exactly the same as $\Phi$, it is nothing more than a criterion for assessing the suitability of $S$ but, depending on the $\pi_i$s, its use is more appropriate in a design--based framework. Anyway, its minimization would again lead to a purposive sample that would not solve the problem of randomization of $S$.\\
A way to recover the use of autocorrelation in a DB sampling strategy, or at least to restore its acceptability, is to consider that in survey design an upper bound to the sampling error is usually imposed to the auxiliary variables rather than on the {\it target} variable. This choice is usually motivated by the practitioners on the basis of the hypothesis that a survey that places its bases on specific levels of precision for a set of auxiliary variables will approximately yield the same sampling errors also for the {\it target} variable. However, in realistic cases of practical interest, this assumption is unlikely to occur and appreciable differences are often measured among the covariates and $Y\bigl(i\bigr)$. In such situations the suggested design could be incorrect as using $\vec{x}_i$ as a proxy for $Y\bigl(i\bigr)$ it could underestimate the sample size needed to reach a predetermined level of precision.\\
An appealing and widely used alternative is to assume a model that links $Y\bigl(i\bigr)$ to $\vec{x}_i$. The assumption underlying this solution is that it is possible to estimate the unknown parameters of such models from data collected in previous surveys. This concept of anticipated variance (AV - or anticipated MSE if the estimator is biased) was introduced by Isaki and Fuller \cite{b37} and is very useful to overcome the hitch of not being able to introduce, at least in the design of the sample, a stochastic model that can take into account the dependence of spatial units. It is defined as the average of the DB variance of $\hat{t}_{HT,Y}$, the HT estimator of the total of $Y$, under a stochastic model $\xi$. To predict the $Y\bigl(i\bigr)$s we assume that, for each unit $i$ of $U$, a linear model $\xi$ holds, given the known auxiliaries $\vec{x}_{i}$s:
\begin{equation}
\label{f:model}
\left\{\begin{tabular} {l}
$Y\bigl(i\bigr)= \vec{x}_{i}^{t}\mathbf{\beta}+\epsilon_i$
\\ $E_\xi \left ( \epsilon_i \right )=0$
\\ $Var_\xi \left ( \epsilon_i \right )=\sigma_i^2$
\\ $Cov_\xi \left ( \epsilon_i \epsilon_j \right )=\sigma_i\sigma_j\rho_{ij}$
\end{tabular}\right.,
\end{equation}
where $E_\xi$, $Var_\xi$, and $Cov_\xi$ denote respectively expectation, variance and covariance with respect to the model $\xi$, $\mathbf{\beta}$ is a vector of regression coefficients, $\epsilon_i$ is a random variable with variance $\sigma_i^2$ and $\rho_{ij}$ is its autocorrelation coefficient. The AV of the unbiased HT estimator of the total of $Y$, under model (\ref{f:model}), is:
\begin{equation}
\label{eq:AV}
AV\left ( \hat{t}_{HT,Y} -t\right )=E_S\left [ \left ( \sum_{i\in S}\frac{\vec{x}_{i}}{\pi_i}-\sum_{i\in U}\vec{x}_{i}\right )^T\mathbf{\beta} \right ]+\sum_{i\in U}\sum_{j\in U}\sigma_i\sigma_j\rho_{ij} \frac{\pi_{ij}-\pi_i\pi_j}{\pi_{ij}}.
\end{equation}
The two components of (\ref{eq:AV}) represent respectively the error implicitly introduced in the estimate of the totals of the auxiliary variables and the existing dependence of the population units. It is clear that uncertainty on estimates can be reduced constraining the units selected to respect the average value of the population's covariates and, assuming that autocorrelation coefficient decreases as the distance $d_{ij}$ between the selected units $i$ and $j$ increases, selecting units as far apart as possible. The logic derived from the AV criterion may, however, sound like an excuse to justify the forced introduction of a model $\xi$ within a DB framework. In this regard, it is interesting to note that the authors of this proposal justify the designs derived from it within a MB and not DB framework \cite{b32}: {\it ``Even though we justify the method by using a superpopulation model, the inference is based on the sampling design. The HT estimator will be efficient if the population is close to a realization from the model, but the estimator maintains desirable properties like design unbiasedness and design consistency even if the model is not properly specified''}.\\
Another motivation for {\it spatially balanced} samples, perhaps less forced and even less arbitrary, has been recently proposed \cite{b05} on the basis of the so called {\it decomposition lemma}, a useful though not so well-known result in sampling theory, which states that \cite[p.~87]{b42}:
\begin{equation}
   \sigma_{\breve{\vec{y}}}^2=\textrm{V}_{S}\left(\bar{\breve{\vec{y}}}_S\right)+\frac{n-1}{n}\textrm{E}_S \left( s_{\breve{\vec{y}},S}^2 \right),
\label{eq:declem}
\end{equation}
where $\breve{\vec{y}}$ is a vector of the expanded-values of the target variable $\vec{y}$ whose generic element is $\breve{y}_i=\pi_i^{-1}y_i$, $\sigma_{\breve{\vec{y}}}^2$ is the constant and unknown population variance of the variable $\breve{\vec{y}}$, $\textrm{V}_{S}\left(\bar{\breve{\vec{y}}}_S\right)$ is the variance between samples of the HT estimator of the mean $\bar{\breve{\vec{y}}}_S=\frac{1}{n}\sum_{i\in{S}}\breve{y}_i$ and $\textrm{E}_S \left( s_{\breve{y},S}^2 \right)$ is the expectation of $ s_{\breve{y},S}^2 = \frac{1}{(n-1)}\sum_{i\in{S}}(\breve{y}_i-\bar{\breve{y}}_S)^2$, i.e. the within sample variance according to the design $P(S)$ (for details see \cite[ch.3]{b42}).\\
It can be seen from (\ref{eq:declem}) that a gain in the efficiency of the HT estimator can be realized either by setting the ${\pi_i}$s in such a way that $\breve{y}$ is approximately constant \cite[p.~53]{b20} or by defining a design $P(S)$ that increases the expected within sample variance or both. Provided that we are dealing with samples $S$ with fixed and known $\pi_i$s, the intuitive explanation for this is that, being constant the unknown variance of the population, the only way to reduce the first term in the right side of (\ref{eq:declem}) is to increase the second term. Thus, we should set the probability $P(S)$ to select a sample $S$ proportional, or even more than proportional, to $s_{\breve{y},S}^2$.\\
Unfortunately, this proposal risks to remain purely theoretical as this parameter is unknown being relative to the unobserved target variable $\vec{y}$. In the spatial interpolation literature \cite{b12,b21}, it is often adopted the assumption that the distance $\mathbf{D}_U$ is highly related to the variance of a variable observed on a set of geo--referenced units. The variogram (or semi variogram), and its shape, is a classical tool to choose how and to what extent the variance of $\breve{y}$ is a function of the distance between the statistical units. Thus, when dealing with spatially distributed populations a promising candidate to play the role of the proxy for $s_{\breve{y},S}^2$ is $\mathbf{D}_U$. The intuitive requirement that a sample should be well--spread over a study region is explainable if and only if there are reasons to assume that $s_{\breve{y},S}^2$ is a monotone increasing function of $\mathbf{D}_U$. This will surely happen when $\breve{y}$ has a linear or monotone spatial trend or when there is spatial dependence. It follows that the distance matrix $\mathbf{D}_S$ between sample units, if appropriately synthesized through an index $\rm{M}\bigl(\mathbf{D}_S\bigr)$, could be an additional criterion to play the same role as SBI and $\Phi$ and is probably much simpler and faster to evaluate.\\
Finally, noting that the Sen-Yates-Grundy (SYG) formulation of the estimator of the sample variance:
\begin{equation}
 \hat{\textrm{V}}_{SYG}\left(\hat{t}_y\right)=-\frac{1}{2}\sum\limits_{i \in S} {\sum\limits_{j  \in S} { \left(\frac{\pi_{ij}-\pi_i\pi_j}{\pi_{ij}}\right) \left(\breve{y}_i-\breve{y}_j\right)^2} },
\label{eq:sygvar}
 \end{equation}
is very similar to the definition of the variogram, another interesting motivation to the practice to spread the sample can be derived from the assumption that, increasing the distance $d_{ij}$ between two units $i$ and $j$, the difference $\left(\breve{y}_i-\breve{y}_j\right)^2$ between the values of the survey variable is expected to increase. As a result an efficient design, to weigh less the expected most relevant differences $\left(\breve{y}_i-\breve{y}_j\right)^2$, should use a set of $\pi_{ij}$ strictly and positively related with $d_{ij}$.
\section{Algorithms for random selection of spatially balanced samples}
\label{s:hpwd}
\begin{figure*}
\centerline{\includegraphics[width=4.8in]{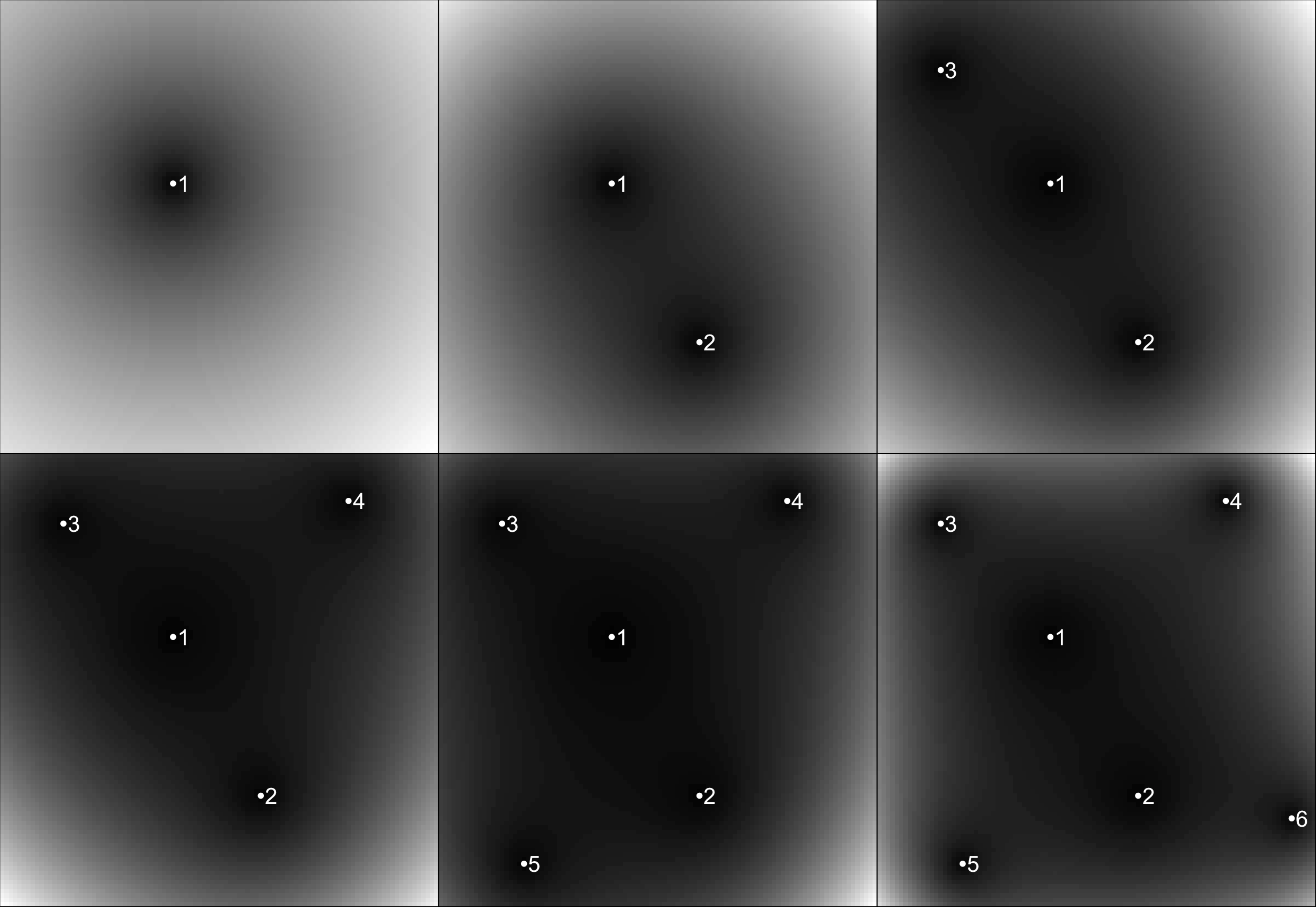}}
\caption{First order inclusion probabilities $\pi_j^t$ of the HPWD of a population on a $100 \times 100$ regolar grid after the selection of the first $t=1,\dots,6$ units with the suggested algorithm (darker is lower and lighter is higher).} 
\label{f:exsel}
\end{figure*}

\begin{figure*}
\centerline{\includegraphics[width=4.8in]{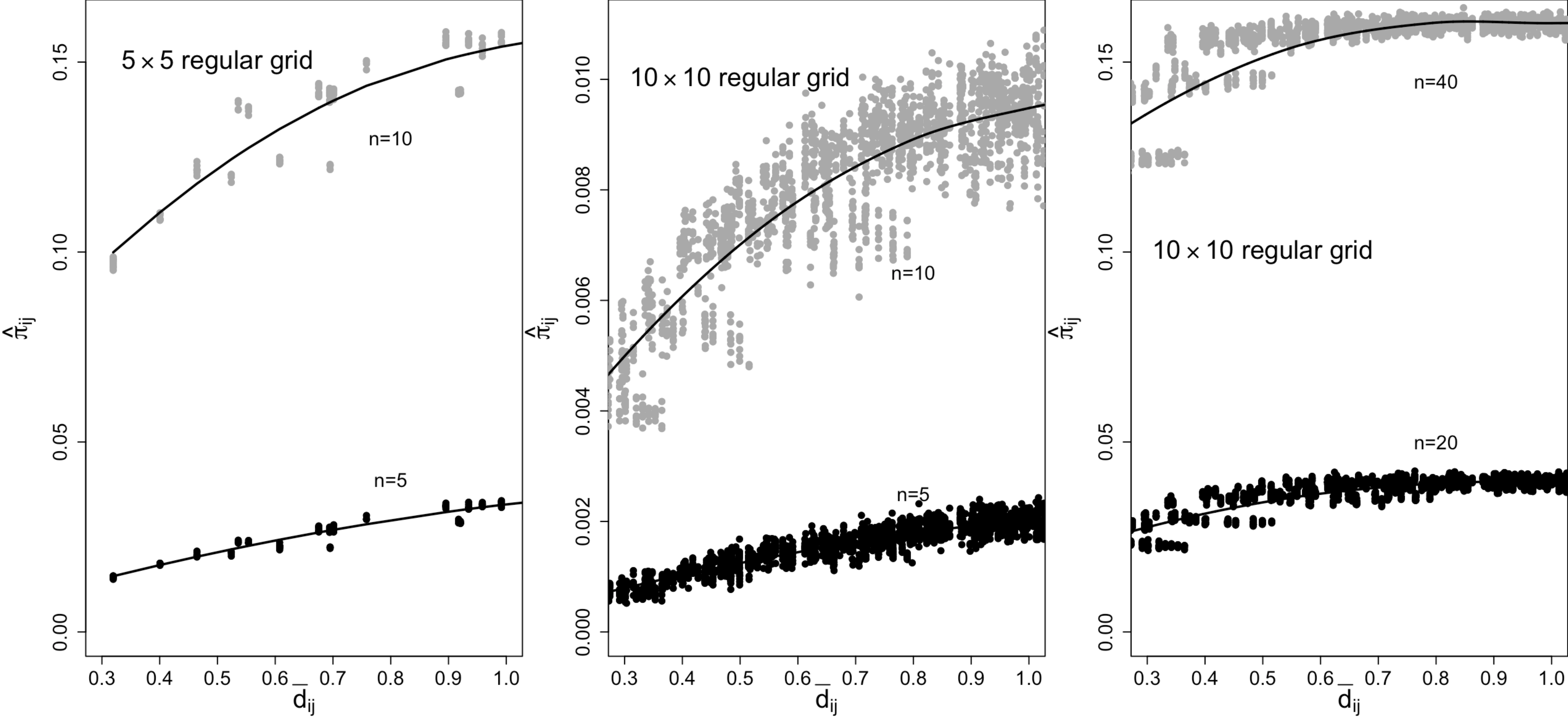}}
\caption{Second order inclusion probabilities $\hat{\pi}_{ij}$ of the HPWD estimated in 100,000 replicated samples in the $5 \times 5$ and $10 \times 10$ regular grid populations for different sample sizes and $\gamma=1$.} 
\label{f:pij}
\end{figure*}

The need to avoid as much as possible to include in the same sample contiguous or too close spatial units is very much felt in real periodic surveys such as those used to estimate land use and land cover \cite[ch. 2]{b06} or to implement a forestry inventory \cite{b32a,b44a}. The strategies adopted in planning these surveys invariably invoke simple and consolidated methods of sampling theory that have the drawback of not being specifically implemented to solve this problem. These solutions often involve systematic selection with a predetermined step and a random starting point, partitioning the area in exactly $n$ contiguous strata from which randomly pick up only one unit per stratum and multi-stage samples with primary units represented by spatial aggregates. Each of these proposals has proved not to be appropriate to capture the potential efficiencies deriving from the existence of spatial trends or dependence. The main reason for these difficulties lies in the evidence that, if the population does not lie on a regular grilling, it is almost impossible to define a sampling step in the absence of a unit order and it is highly subjective to set up a partition of the region in $n$ parts. The choices made to overcome these shortcomings seriously affect the results.\\
One of the first attempts to formalize the problem, increasing the amount of information collected by avoiding the selection of pairs of contiguous units, was done in the pioneering work \cite{b36} that, however, explicitly requires the introduction of an exogenous ordering of the population units. The line drawn from this preliminary work has remarkably influenced the approaches adopted in the following studies on this topic. Many algorithms, even recently proposed in the literature \cite{b55}, are primarily based on the search for a one-dimensional sorting of multidimensional units by studying the best mapping of $\mathbb{R}^d$ in $\mathbb{R}$, while trying to preserve some multidimensional features of the population, and then use this induced ordering to systematically select the sample units. The fundamental principle is to extend the use of systematic sampling to two or more dimensions, even when the population is not a regular grid. Developed by the Environmental Protection Agency (EPA) and widely used in most of its surveys, the Generalized Random Tessellation Stratified (GRTS) design \cite{b55} is mainly based on such a logic. A multilevel grid hierarchy is used to index the units through a tree structure.\\
The two parts of the AV (\ref{eq:AV}) suggested instead the use of two different classes of methods, the first derived from the restriction of the sample space considered acceptable and the second from the assumption that dependence between units decreases as their distance increases.
Despite the assonance of the two names, the notion of {spatially balanced samples} is quite far from that of {\it balanced samples} used to denote those samples that respect the known totals of a set of auxiliary variables, their selection is made at random on a restricted support $\bar{\mathcal{S}} \subset \mathcal{S}$ of all the possible samples \cite{b57,b58,b60}. Thus, for given $\pi_i$s the set of {\it balanced samples} are defined as all those samples $S$ for which it occurs $\lvert \hat{t}_{HT,\vec{x}_c}-t_{HT,\vec{x}_c} \rvert \leq \omega_b, \: \forall \: c=\{1,\dots,C\}$, where $\omega_b$ is an acceptable upper bound. These restrictions represent the intuitive requirement that the sample estimates of the total of a covariate should be as close as possible to the known total of the population. In a spatial context, this constraint could be applied by imposing that, for any $S$, the first $p$ moments of each coordinate should coincide with the first $p$ moments of the population, implicitly assuming that $Y\bigl(i\bigr)$ follows a polynomial spatial trend of order $p$. This logic was subsequently extended to approximate any nonlinear trends through penalized splines with particular reference to the space \cite{b09}.\\
The CUBE algorithm \cite{b11,b18}, proposed as a general (not spatial) solution, is the tool used to achieve these samples. In recent studies \cite{b05,b06}, these sampling designs have demonstrated that they can effectively exploit the presence of linear and nonlinear trends but are often unable to capture the presence of dependence or homogeneity in $Y\bigl(i\bigr)$.\\
The second class is instead more articulated and consisting of algorithms of different nature. A selection strategy conceived with the clear goal of minimizing (\ref{f:indexsb}) should use $\mathbf{D}_U$ as the only known information available on the spatial distribution of the sample units. Its use implies the adoption of the already mentioned intuitive criterion that units that are close should seldom appear simultaneously in the sample.
Recalling (\ref{eq:sygvar}), this request can be claimed as reasonable if we expect that, increasing $d_{ij}$, the difference $\left(\breve{y}_i-\breve{y}_j\right)^2$ always increases. In such a situation, it is clear that the variance of the HT estimator will necessarily decrease if we set high joint inclusion probabilities to couples with very different values of the {\it target} variable as they are far each other. Following an approach based on distances, inspired by purely MB assumptions on the dependence of the stochastic process generating the data, \cite{b00} suggested a sampling strategy: the dependent areal units sequential technique (DUST). Starting with a unit selected at random, say $i$, at every step $t<n$, the selection probabilities are updated according to a multiplicative rule depending by a tuning parameter useful to control the distribution of the sample over the study region. This algorithm, or at least the design that it implies, can be easily interpreted and analyzed in a DB perspective in particular referring to a careful estimation and analysis of its first and second order inclusion probabilities.\\
Another solution, based on a classic list sequential algorithm \cite{b57}, was suggested by \cite{b29}. Introduced as a variant of the correlated Poisson sampling, in the SCPS (Spatially Correlated Poisson Sampling) for each unit, at every step $t$, it updates the inclusion probabilities according to a rule in such a way that the required inclusion probabilities are respected. The suggested {\it maximal weights} criterion, used to update the inclusion probabilities at each step, provides as much weight as possible to the closest unit, then to the second closest unit and so on. The resulting sample is then obtained in $n \leq t \leq N$ steps depending on the randomness according to which the last unit is added.\\
A procedure to select samples with fixed $\pi_i$ and correlated $\pi_{ij}$ was derived by \cite{b31} as an extension of the pivotal method initially introduced to select $\pi$ps samples \cite{b17}. It is essentially based on an updating rule of the probabilities that at each step should locally keep the sum of the updated probabilities as constant as possible and differ from each other in a way not to choose the two nearby units. This method is referred to as the local pivotal method (LPM) and obtain a sample in, again, $n \leq t \leq N$ steps. This solution has been also integrated with the {\it balancing} property, typical of the CUBE algorithm, to simultaneously minimize both the two members of the AV (\ref{eq:AV}) \cite{b32}. A comprehensive review of the main {\it spatially balanced} samples selection methods can be found in \cite{b07}.\\
According to (\ref{eq:declem}) the attention should be moved to the definition of a design with probability proportional to some synthetic index $\rm{M}\bigl( \cdot \bigr)$ of the within sample distance matrix when it is observed within each possible sample: $P(S)=\rm{M}\bigl(\mathbf{D}_S\bigr)/\sum_{S \in \mathcal{S}} \rm{M}\bigl(\mathbf{D}_S\bigr)$ \cite{b05}. This innovative proposal is quite dangerous because extracting realizations from the complete $P(S)$ could make it very difficult to control or specify the $\pi_i$s and the $\pi_{ij}$s whose knowledge is required by the HT estimator to be able to yield total and variance estimates. Despite this initial difficulty it may open a very interesting window on how to use indices, such as $\Phi$ or SBI, within a DB framework. Select the whole sample with probability proportional to the index instead of maximizing or minimizing the index itself could remove all the doubts that sometimes arise from choosing the units with certainty. Interpreted in this way, this proposal could thus represent a bridge between the MB and DB frameworks to setting up a spatial sampling strategy.\\
Among the possible summary indexes $\rm{M}\bigl( \cdot \bigr)$ of the distance matrix, the most promising results were provided by ${\rm{M}} \left( {{\bf{D}}_S} \right) = \prod\limits_{i \in S} {\prod\limits_{j \ne i;j \in S} {d_{ij}^\gamma}}$ introducing a design proportional to the products of the within sample distance matrix (PWD) that depends on $\gamma$, a tuning parameter that can be used without limits to increase or decrease the effects of the distance matrix on the spread of the sample over the study region. In \cite{b05} it is suggested to start with a SRS without replacement and repeatedly and randomly exchanging a unit included in the sample with a unit not included in the sample with probability equal to the exponential of the ratio between the two indexes before and after the exchange. This is an MCMC iterative procedure that for a suitable choice of the number of iterations will generate a random outcome from the multivariate distribution $P(S)$.\\
These procedures are quick and efficient and thus have a practical applicability in real spatial surveys if the size $N$ of $U$ is not prohibitive. The main reason is that they strictly depend on $N$ in the number of attempts or steps needed to select a sample. According to the definition \cite[p. 35]{b57} {\it ``A sampling design of fixed sample size n is said to be draw--by--draw if, at each one of the $n$ steps of the procedure, a unit is definitively selected in the sample''}, a possible way to reduce the computational burden is to look for a draw--by--draw alternative to the listed methods. With regard to the PWD, deriving its probabilities from the product of the distance between units, a natural candidate could arise from the idea of iteratively updating, through a product, the $\pi_i$s at each step.\\
The algorithm suggested in this paper, starts by randomly selecting a unit $i$ with equal probability. Then, at every step $t \leq n$, the algorithm updates the selection probabilities $\pi_j^t$s of every other unit $j$ of the population according to the rule (see Figure \ref{f:exsel}):
\begin{equation}
\pi_j^t=\dfrac{\pi_j^{t-1}\bar{d}_{ij}}{\sum_{j \in U}\pi_j^{t-1}\bar{d}_{ij}} \quad \forall j \in U,
\label{eq:cfedi}
\end{equation}
where $\bar{d}_{ij}=\Psi \bigl( d_{ij}^\gamma \bigr)$ is an appropriate transformation applied to $\mathbf{D}_U$ in order to standardize it so that it will have known and fixed products by row $\prod_{i\neq j,i \in U} d_{ij}$ and column $\prod_{i\neq j,j \in U} d_{ij}$. Notice that, as $\bar{d}_{ii}=0$ by the definition of distance, $\pi_i^t$ will be necessarily equal to 0 for all units $i$ already selected in the sample in previous steps, thus preventing random selection with replacement. Criterion (\ref{eq:cfedi}) can be basically considered as a heuristic method (HPWD) to generate samples approximately with the same probabilities of the PWD but with a much smaller number of steps. From a practical point of view, the idea behind (\ref{eq:cfedi}) is very similar to the DUST method already suggested in the literature \cite{b00} although with two substantive differences: (\ref{eq:cfedi}) is motivated by solid theoretical bases while the DUST did not find any justifications either according to a MB or DB logic and in the HPWD it is possible to control the $\pi_i$s while with DUST this is impossible with the result that the HT estimator systematically yields irreparably biased estimates.\\
In \cite{b05} it is advised that the $\pi_i$s and the $\pi_{ij}$s of the PWD follow the rules:
\begin{equation}
\hat\pi _i  = k_1 \left( {\prod_{j  = 1}^N {d_{ij} } } \right)^{k_2},
\label{eq:pincpiprod}
\end{equation}
\begin{equation}
\hat\pi _{ij}  = k_3 \left( {\prod_{i  = 1}^N {d_{ij} } \times \prod_{j  = 1}^N {d_{ij} } } \right)^{k_4 } \left( {d_{ij} } \right)^{k_5},
\label{eq:pincpijprod}
\end{equation}
where $k_1$, $k_2$, $k_3$, $k_4$ and $k_5$ are parameters mainly depending on the sample and population sizes. Notice that, to ensure the symmetry of the $\pi _{ij}$s, the rows and columns marginal products of the distance matrix have the same parameter.\\
From (\ref{eq:pincpiprod}) it is clear that if we want to fix the $\pi_i$s, it is enough to standardize the $d_{ij}$s using the $\bar{d}_{ij}$s specifying an appropriate standardizing function $\Psi$. For this role \cite{b05} suggest to iteratively constrain to known totals, the rows (or columns) sums of the logarithmic transformed matrix. When the known totals are all equal to a constant, this is known to be a very simple and accurate method to scale a symmetric matrix to a doubly stochastic matrix \cite{b39}.\\
To verify if (\ref{eq:pincpiprod}) and (\ref{eq:pincpijprod}) can be used as working rules to specify the $\pi_i$s we can generate as many independent replicates from HPWD as needed and the $\pi_i$s and the $\pi_{ij}$s may be estimated by the frequency with which a unit or a pair of units are selected. These $\hat{\pi}_i$s and $\hat{\pi}_{ij}$s can both be adopted in the estimation process instead of their theoretical counterparts \cite{b22,b23} and also modeled to verify the fit of (\ref{eq:pincpijprod}).\\
To this purpose we selected 100,000 replicated samples of size $n=\{5,10\}$ from a $5 \times 5$ regular grid and of size $n=\{5,10,20,40\}$ from a $10 \times 10$ regular grid by using the HPWD design with 3 different values of the parameter $\gamma$=\{1, 5, 10\}.\\
As far as the empirical frequencies of units in selected samples are concerned, in Table (\ref{t:tabpigrid}) are shown the values of the index:
\begin{equation}
\label{eq:cvpi}
CV\left ( \hat{\pi}_i \right )=\dfrac{N}{n}\sqrt{\dfrac{\sum_{i \in U}\left (  \hat{\pi}_i-\dfrac{n}{N}\right )^2}{N}}\times 100.
\end{equation}
From the results in Table (\ref{t:tabpigrid}) we can observe that the $\pi_i$s are approximately constant once the row (and column) products of the $\bar{d}_{ij}$s are fixed as constant with an error that increases with $\gamma$ and decreases with the increase of $n$. The first effect is probably due to the fact that increasing $\gamma$ necessarily increases the concentration of the $\bar{d}_{ij}$s in few very high values and many low values, which undoubtedly reduces the accuracy of the matrix standardization and of the logarithmic transformation. While the second aspect occurs as increasing the number of random selections the empirical $\hat{\pi}_i$s are closer to the theoretical $\pi_i$s.\\
\begin{table*}
\caption{Relative efficiency of the estimated first order inclusion probabilities $CV\left ( \hat{\pi}_i \right )$ (\ref{eq:cvpi}) of the HPWD estimated in 100,000 replicated samples in the $5 \times 5$ and $10 \times 10$ regular grid populations for different sample sizes and $\gamma$.}
\label{t:tabpigrid}
\begin{center}
\setlength\tabcolsep{3 pt}
\begin{tabular}{r|rr|rrrr}
 & \multicolumn{2}{c|}{{$5 \times 5$}} & \multicolumn{4}{c}{{$10 \times 10$}} \\
$\gamma$ & $n=5$ & $n=10$ & $n=5$ & $n=10$ & $n=20$ & $n=40$ \\
 \hline
 1 & 1.852 & 0.653 & 2.864 & 2.688 & 1.477 & 0.453 \\ 
 5 & 8.543 & 2.457 & 13.373 & 8.082 & 3.659 & 1.469 \\ 
10 & 12.722 & 5.297 & 21.358 & 13.131 & 6.080 & 2.583 \\ 
 \hline
\end{tabular}
\end{center}
\end{table*}
With regard to the $\hat{\pi}_{ij}$s the evidence from Figure (\ref{f:pij}) and Table (\ref{t:tabpijgrid}) is that (\ref{eq:pincpijprod}) fits enough well supporting the assumption of such a relationship between the distance and the estimated second order probabilities. However, some trends can be noticed not only in $R^2$ but also in the intercept ($\hat{k}_3$) and in the slope ($\hat{k}_5$) estimated through ordinary least squares (OLS) on a logarithmic transformation of (\ref{eq:pincpijprod}). As expected increasing $\gamma$ the effects of the distance on the $\hat{\pi}_{ij}$s are more sensible leading to a more scattered spatial distribution of the sampling units. An evidence that is mitigated by the increase of $n$ as in this case the HPWD has less space to better spread the units.\\
\begin{table*}
\caption{Regression parameters of the model (\ref{eq:pincpijprod}) for the $\hat{\pi}_{ij}$ of the HPWD estimated in 100,000 replicated samples in the $5 \times 5$ and $10 \times 10$ regular grid populations for different sample sizes and $\gamma$.}
\label{t:tabpijgrid}
\begin{center}
\setlength\tabcolsep{3 pt}
\begin{tabular}{c|r|r|rrr}
population grid & $\gamma$ & $n$ & \multicolumn{1}{c}{{$log \left (\hat{k}_3 \right )$}} & \multicolumn{1}{c}{{$\hat{k}_5$}} & \multicolumn{1}{c}{{$R^2$}} \\
 \hline
$5 \times 5$ & 1 & 5 & $-3.377 (0.006)^{***}$ & $0.718 (0.012)^{***}$ & 0.943 \\
$5 \times 5$ & 5 & 5 & $-3.025 (0.045)^{***}$ & $3.022 (0.085)^{***}$ & 0.846 \\
$5 \times 5$ & 10 & 5 & $-2.668 (0.096)^{***}$ & $5.563 (0.180)^{***}$ & 0.806 \\
$5 \times 5$ & 1 & 10 & $-1.842 (0.005)^{***}$ & $0.388 (0.010)^{***}$ & 0.872 \\
$5 \times 5$ & 5 & 10 & $-1.606 (0.027)^{***}$ & $1.205 (0.051)^{***}$ & 0.709 \\
$5 \times 5$ & 10 & 10 & $-1.426 (0.048)^{***}$ & $1.918 (0.091)^{***}$ & 0.660 \\
 \hline
$10 \times 10$ & 1 & 5 & $-6.149 (0.003)^{***}$ & $0.831 (0.004)^{***}$ & 0.917 \\
$10 \times 10$ & 5 & 5 & $-4.951 (0.052)^{***}$ & $5.799 (0.073)^{***}$ & 0.830 \\
$10 \times 10$ & 10 & 5 & $-4.116 (0.074)^{***}$ & $11.453 (0.105)^{***}$ & 0.763 \\
$10 \times 10$ & 1 & 10 & $-4.587 (0.003)^{***}$ & $0.598 (0.004)^{***}$ & 0.862 \\
$10 \times 10$ & 5 & 10 & $-4.066 (0.014)^{***}$ & $2.235 (0.020)^{***}$ & 0.774 \\
$10 \times 10$ & 10 & 10 & $-3.039 (0.052)^{***}$ & $5.485 (0.074)^{***}$ & 0.599 \\
$10 \times 10$ & 1 & 20 & $-3.157 (0.002)^{***}$ & $0.361 (0.003)^{***}$ & 0.782 \\
$10 \times 10$ & 5 & 20 & $-2.916 (0.010)^{***}$ & $0.988 (0.014)^{***}$ & 0.591 \\
$10 \times 10$ & 10 & 20 & $-2.714 (0.017)^{***}$ & $1.607 (0.024)^{***}$ & 0.540 \\
$10 \times 10$ & 1 & 40 & $-1.789 (0.001)^{***}$ & $0.165 (0.002)^{***}$ & 0.722 \\
$10 \times 10$ & 5 & 40 & $-1.725 (0.004)^{***}$ & $0.316 (0.006)^{***}$ & 0.406 \\
$10 \times 10$ & 10 & 40 & $-1.685 (0.007)^{***}$ & $0.430 (0.010)^{***}$ & 0.337 \\
\hline
\multicolumn{6}{l}{\scriptsize{$^{***}p<0.001$, $^{**}p<0.01$, $^*p<0.05$}}
\end{tabular}
\end{center}
\end{table*}
Regarding the evaluation of the $\pi _{ij}$s we should also consider that their use is expected only in HT variance estimation and that a spatial design producing well spread samples necessarily will have very small joint inclusion probabilities for nearby units. Even though HPWD has the advantage to produce $\pi _{ij}$s that can be shown to be always strictly positive, although it is still unbiased, the HT variance estimator could become extremely unstable if there is some very small $\pi _{ij}$. Thus, using HPWD, the HT estimation of variance is always possible, and this is a clear advantage, but it cannot always be recommended as a good solution. In such circumstances it could be better the use of alternatives variance estimators as a MB approach to the problem \cite{b03} or the local variance estimator suggested by \cite{b54} that represents a general solution as it does not need the $\pi _{ij}$s.
\section{Sampling Designs Comparison on Artificial and Real Populations}
\label{s:compar}
\begin{figure*}
\centerline{\includegraphics[width=4.8in]{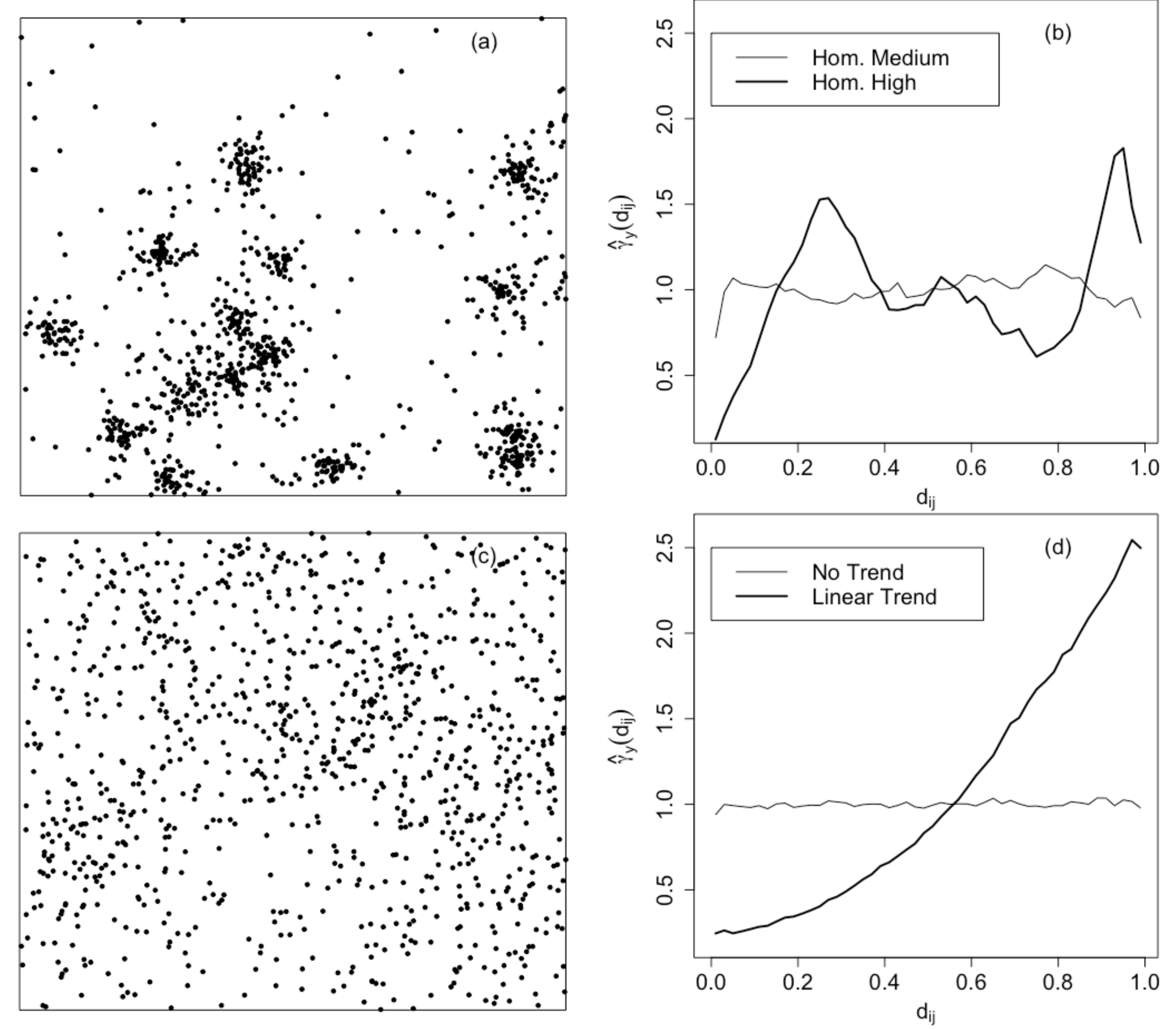}}
\caption{Spatial distribution of the simulated populations: (a) clustered and (c) sparse. The variograms are reported for Medium and High autocorrelation without trend (b) and for No Trend and Linear Trend without autocorrelation (d).} 
\label{f:popsim}
\end{figure*}
\begin{figure*}
\centerline{\includegraphics[width=4.8in]{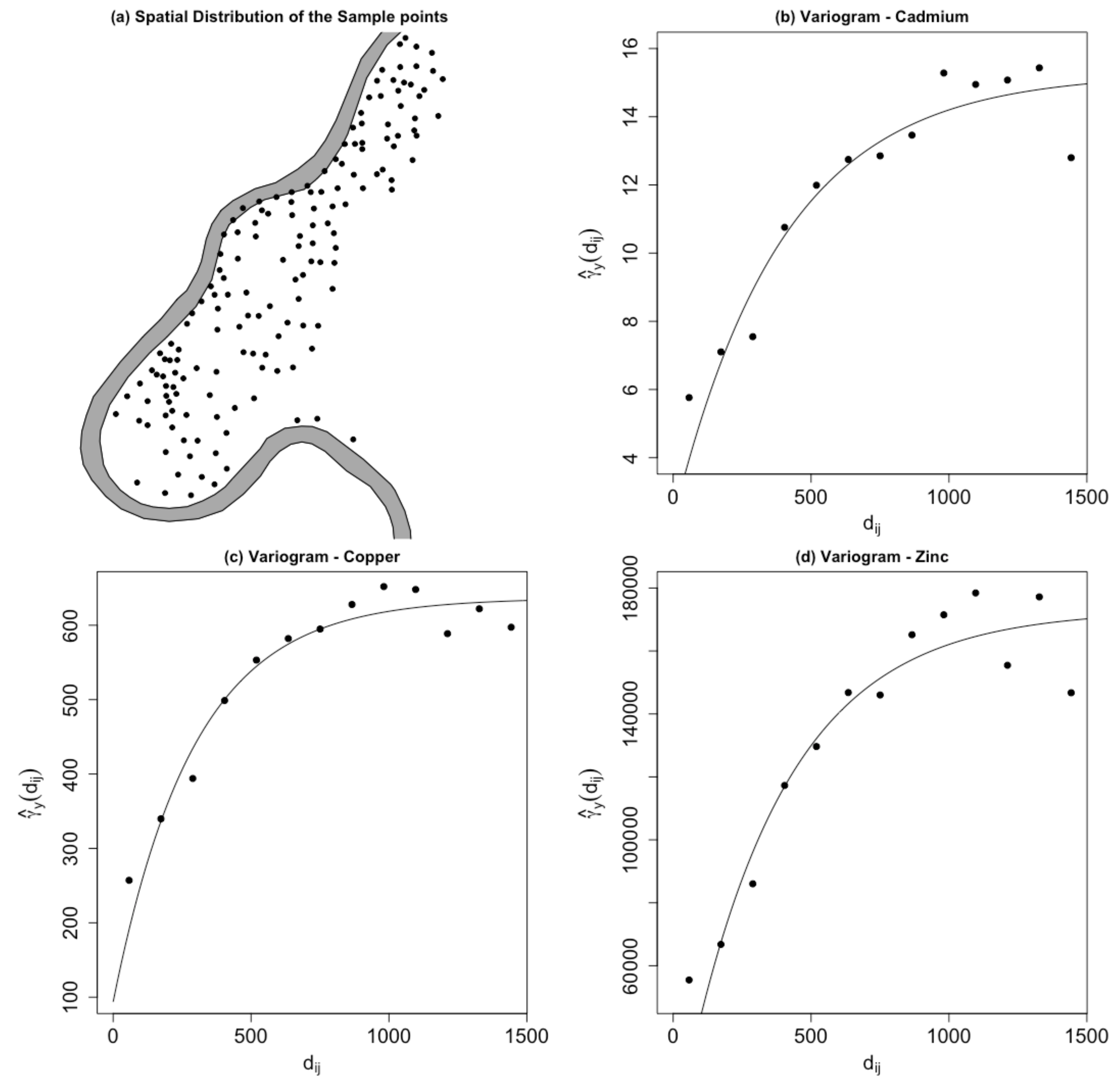}}
\caption{Spatial distribution of the Meuse river data (a). The fitted exponential variogram models are reported for the variables Cadmium (b), Copper (c) and Zinc (d).} 
\label{f:meuse}
\end{figure*}
\begin{figure*}
\centerline{\includegraphics[width=4.8in]{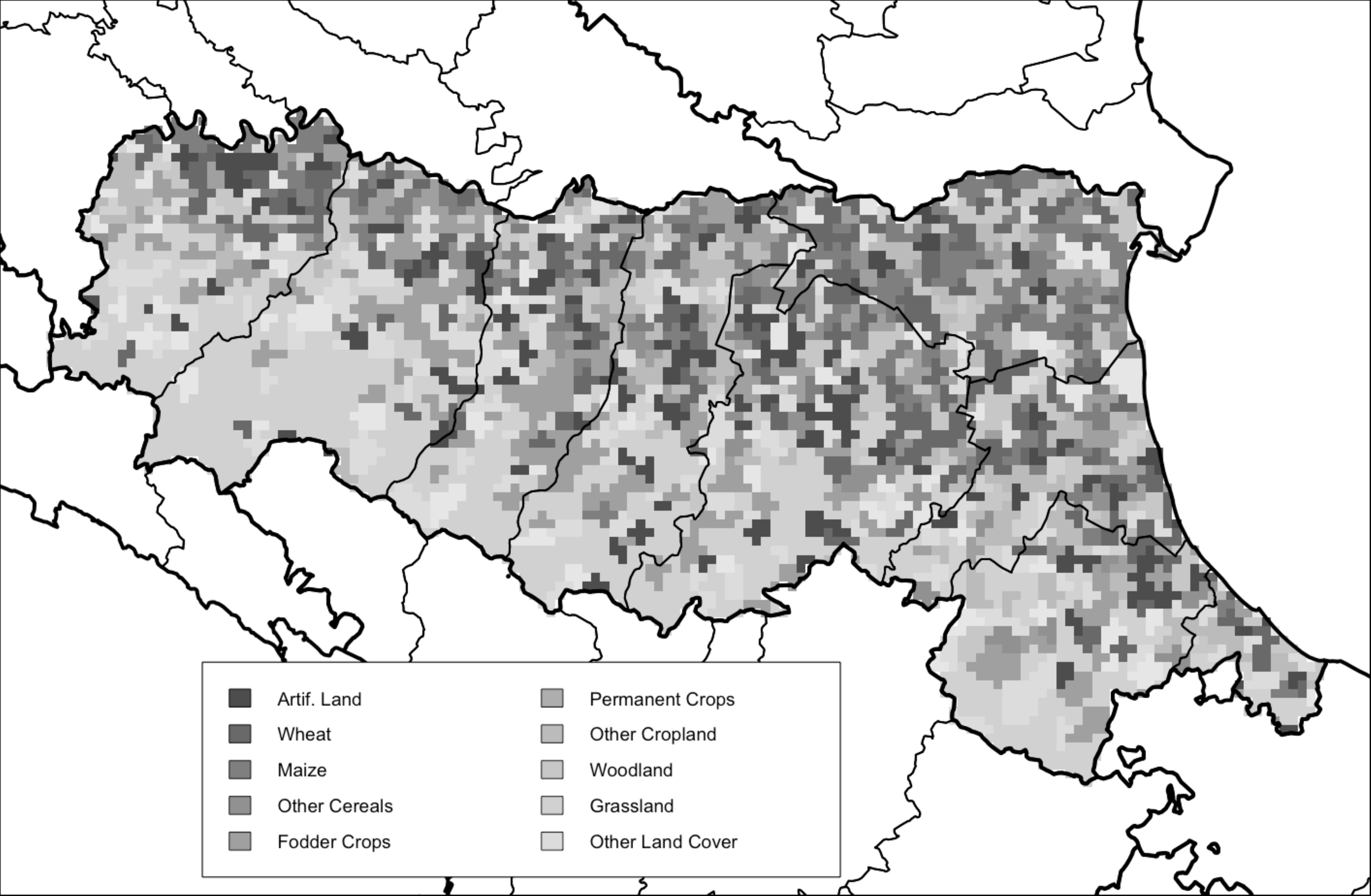}}
\caption{Land Cover LUCAS data on a $2km \times 2km$ population grid in the Italian region Emilia Romagna overlaid on the NUTS2 and NUTS3 boundaries (wider and narrower lines respectively).} 
\label{f:lucas}
\end{figure*}
\begin{table*}
\caption{Relative efficiency of the estimated first order inclusion probabilities $CV\left ( \hat{\pi}_i \right )$ (\ref{eq:cvpi}) of the HPWD in 10,000 replicated samples in the LUCAS - Emilia Romagna population, Meuse river, clustered and sparse simulated populations for different sample sizes and $\gamma$.}
\label{t:tabpi}
\begin{center}
\setlength\tabcolsep{1.5 pt}
\begin{tabular}{r|rrr|rr|rrr|rrr}
 & \multicolumn{3}{c|}{{LUCAS - Emilia Romagna}} &\multicolumn{2}{c|}{{Meuse river}} & \multicolumn{3}{c|}{{Sim. Clustered Population}} & \multicolumn{3}{c}{{Sim. Sparse Population}}\\
$\gamma$ & $n=100$ & $n=300$ & $n=600$ & $n=10$ & $n=50$ & $n=50$ & $n=100$ & $n=200$ & $n=50$ & $n=100$ & $n=200$\\
 \hline
1 & 7.658 & 4.480 & 3.001 & 5.695 & 2.271 & 10.542 & 7.616 & 5.788 & 5.766 & 5.396 & 5.153 \\ 
5 & 11.183 & 7.505 & 5.519 & 18.393 & 12.329 & 28.539 & 28.146 & 26.747 & 21.514 & 26.076 & 26.237 \\ 
10 & 16.264 & 11.682 & 8.864 & 32.002 & 20.952 & 43.558 & 47.529 & 40.981 & 37.762 & 44.454 & 41.986 \\ 
 \hline
\end{tabular}
\end{center}
\end{table*}
\begin{table*}
\caption{Relative efficiency of the sample mean $\left(\frac{RMSE}{RMSE_{SRS}}\right)$ and average $SBI$ for each design estimated in 10,000 replicated samples in the clustered and sparse populations for different sample sizes, trend and autocorrelation.}
\label{t:tabsim}
\begin{center}
\setlength\tabcolsep{3 pt}
\begin{tabular}{l|c|ccccc|ccccc}
 & & \multicolumn{5}{c|}{{Clustered Population}} & \multicolumn{5}{c}{{Sparse Population}} \\ [1pt]
 & & \multicolumn{2}{c}{{No Trend}} & \multicolumn{2}{c}{{Linear Trend}} & & \multicolumn{2}{c}{{No Trend}} & \multicolumn{2}{c}{{Linear Trend}} &\\ [1pt]
 & & \multicolumn{2}{c}{{Autocorrelation}} & \multicolumn{2}{c}{{Autocorrelation}} & Average & \multicolumn{2}{c}{{Autocorrelation}} & \multicolumn{2}{c}{{Autocorrelation}} & Average \\ [1pt]
{Design}  & {$n$} & {Medium} & {High} & {Medium} & {High} & $SBI$ & {Medium} & {High} & {Medium} & {High} & $SBI$\\
 \hline
GRTS & 50 & 0.963 & 0.612 & 0.519 & 0.349 & 0.308 & 0.990 & 0.777 & 0.549 & 0.520 & 0.132 \\
LPM & 50 & 0.935 & 0.591 & 0.498 & 0.322 & 0.265 & 0.991 & 0.698 & 0.522 & 0.449 & 0.094 \\
SCPS & 50 & 0.935 & 0.592 & 0.498 & 0.336 & 0.279 & 0.999 & 0.687 & 0.515 & 0.437 & 0.093 \\
PWD1 & 50 & 0.946 & 0.624 & 0.515 & 0.359 & 0.272 & 1.002 & 0.774 & 0.542 & 0.509 & 0.136 \\
PWD5 & 50 & 0.957 & 0.491 & 0.460 & 0.246 & 0.135 & 1.004 & 0.619 & 0.503 & 0.402 & 0.065 \\
PWD10 & 50 & 0.975 & 0.466 & 0.443 & 0.221 & 0.104 & 0.991 & 0.566 & 0.487 & 0.385 & 0.050 \\
HPWD1 & 50 & 0.951 & 0.611 & 0.507 & 0.339 & 0.269 & 1.003 & 0.760 & 0.524 & 0.478 & 0.123 \\
HPWD5 & 50 & 0.935 & 0.504 & 0.470 & 0.262 & 0.166 & 1.005 & 0.633 & 0.502 & 0.393 & 0.067 \\
HPWD10 & 50 & 0.955 & 0.466 & 0.471 & 0.247 & 0.147 & 1.002 & 0.597 & 0.494 & 0.382 & 0.058 \\
 \hline
GRTS & 100 & 0.924 & 0.557 & 0.481 & 0.299 & 0.230 & 0.959 & 0.669 & 0.514 & 0.425 & 0.142 \\
LPM & 100 & 0.904 & 0.521 & 0.462 & 0.280 & 0.191 & 0.965 & 0.629 & 0.501 & 0.385 & 0.105 \\
SCPS & 100 & 0.908 & 0.517 & 0.474 & 0.281 & 0.203 & 0.968 & 0.630 & 0.508 & 0.387 & 0.104 \\
PWD1 & 100 & 0.919 & 0.565 & 0.480 & 0.315 & 0.213 & 0.981 & 0.704 & 0.514 & 0.446 & 0.149 \\
PWD5 & 100 & 0.907 & 0.443 & 0.430 & 0.225 & 0.107 & 0.941 & 0.538 & 0.477 & 0.354 & 0.078 \\
PWD10 & 100 & 0.907 & 0.416 & 0.421 & 0.214 & 0.093 & 0.935 & 0.515 & 0.474 & 0.342 & 0.068 \\
HPWD1 & 100 & 0.925 & 0.542 & 0.467 & 0.293 & 0.205 & 0.973 & 0.674 & 0.510 & 0.417 & 0.136 \\
HPWD5 & 100 & 0.888 & 0.444 & 0.440 & 0.236 & 0.121 & 0.957 & 0.549 & 0.474 & 0.346 & 0.080 \\
HPWD10 & 100 & 0.894 & 0.417 & 0.432 & 0.222 & 0.103 & 0.948 & 0.510 & 0.471 & 0.339 & 0.070 \\
 \hline
GRTS & 200 & 0.900 & 0.505 & 0.454 & 0.253 & 0.199 & 0.947 & 0.613 & 0.491 & 0.387 & 0.169 \\
LPM & 200 & 0.874 & 0.466 & 0.429 & 0.233 & 0.164 & 0.938 & 0.551 & 0.478 & 0.336 & 0.132 \\
SCPS & 200 & 0.877 & 0.459 & 0.425 & 0.234 & 0.172 & 0.941 & 0.550 & 0.473 & 0.333 & 0.135 \\
PWD1 & 200 & 0.895 & 0.512 & 0.448 & 0.263 & 0.202 & 0.939 & 0.618 & 0.483 & 0.389 & 0.179 \\
PWD5 & 200 & 0.829 & 0.395 & 0.400 & 0.197 & 0.123 & 0.889 & 0.473 & 0.450 & 0.285 & 0.109 \\
PWD10 & 200 & 0.826 & 0.389 & 0.401 & 0.191 & 0.118 & 0.887 & 0.469 & 0.452 & 0.280 & 0.104 \\
HPWD1 & 200 & 0.890 & 0.483 & 0.436 & 0.249 & 0.191 & 0.947 & 0.589 & 0.486 & 0.362 & 0.166 \\
HPWD5 & 200 & 0.836 & 0.400 & 0.399 & 0.196 & 0.125 & 0.902 & 0.470 & 0.456 & 0.284 & 0.109 \\
HPWD10 & 200 & 0.810 & 0.386 & 0.385 & 0.182 & 0.112 & 0.864 & 0.443 & 0.452 & 0.268 & 0.097 \\
\hline
\end{tabular}
\end{center}
\end{table*}
\begin{table*}
\caption{Relative efficiency of the sample mean $\left(\frac{RMSE}{RMSE_{SRS}}\right)$ and average $SBI$ for each design estimated in 10,000 replicated samples in the Meuse river population for two different sample sizes.}
\label{t:tabmeuse}
\begin{center}
\setlength\tabcolsep{3 pt}
\begin{tabular}{l|ccccc|ccccc}
 & \multicolumn{5}{c|}{{Sample Size $n=10$}} & \multicolumn{5}{c}{{Sample Size $n=50$}} \\ [1pt]
{Design} & {Cadmium} & {Copper} & {Lead} & {Zinc} & {Av. $SBI$} & {Cadmium} & {Copper} & {Lead} & {Zinc} & {Av. $SBI$}\\ [1pt]
 \hline
GRTS & 0.897 & 0.887 & 0.884 & 0.908 & 0.173 & 0.839 & 0.809 & 0.806 & 0.838 & 0.176 \\
LPM & 0.884 & 0.869 & 0.862 & 0.901 & 0.129 & 0.762 & 0.724 & 0.736 & 0.757 & 0.133 \\
SCPS & 0.883 & 0.864 & 0.853 & 0.893 & 0.120 & 0.755 & 0.708 & 0.705 & 0.735 & 0.138 \\
PWD1 & 0.907 & 0.895 & 0.905 & 0.915 & 0.178 & 0.801 & 0.770 & 0.774 & 0.797 & 0.185 \\
PWD5 & 0.770 & 0.782 & 0.773 & 0.799 & 0.086 & 0.688 & 0.631 & 0.647 & 0.672 & 0.119 \\
PWD10 & 0.704 & 0.747 & 0.727 & 0.747 & 0.059 & 0.664 & 0.602 & 0.619 & 0.646 & 0.110 \\
HPWD1 & 0.902 & 0.888 & 0.919 & 0.915 & 0.181 & 0.798 & 0.760 & 0.775 & 0.805 & 0.189 \\
HPWD5 & 0.767 & 0.782 & 0.784 & 0.797 & 0.085 & 0.678 & 0.621 & 0.648 & 0.671 & 0.125 \\
HPWD10 & 0.722 & 0.754 & 0.735 & 0.754 & 0.058 & 0.671 & 0.609 & 0.626 & 0.649 & 0.112 \\
\hline
\end{tabular}
\end{center}
\end{table*}
\begin{table*}
\caption{Relative efficiency of the Land Cover codes area estimates $\left(\frac{RMSE}{RMSE_{SRS}}\right)$ and average $SBI$ for each design estimated in 10,000 replicated samples in the Emilia-Romagna LUCAS population for different sample sizes.}
\label{t:tablucas}
\begin{center}
\setlength\tabcolsep{2 pt}
\begin{tabular}{l|c|cccccccccc|c}
 &  & Artif. & Wheat & Maize & Other & Fodder & Perman. & Other & Woodl. & Grassl. & Ot. Land & Average \\
{Design}  & {$n$} & Land &  &  & Cereals & Crops & Crops & Cropl. &  &  & Cover & $SBI$\\
 \hline
GRTS & 100 & 0.981 & 0.950 & 0.933 & 0.981 & 0.937 & 0.939 & 0.961 & 0.819 & 0.967 & 0.976 & 0.288 \\
LPM & 100 & 0.956 & 0.947 & 0.935 & 0.968 & 0.937 & 0.903 & 0.954 & 0.801 & 0.950 & 0.942 & 0.110 \\
SCPS & 100 & 0.958 & 0.937 & 0.949 & 0.973 & 0.921 & 0.916 & 0.941 & 0.796 & 0.954 & 0.955 & 0.070 \\
PWD1 & 100 & 0.965 & 0.952 & 0.936 & 0.965 & 0.942 & 0.936 & 0.963 & 0.825 & 0.965 & 0.954 & 0.054 \\
PWD5 & 100 & 0.956 & 0.943 & 0.922 & 0.968 & 0.917 & 0.915 & 0.942 & 0.796 & 0.953 & 0.953 & 0.076 \\
PWD10 & 100 & 0.951 & 0.935 & 0.939 & 0.961 & 0.925 & 0.911 & 0.954 & 0.793 & 0.963 & 0.935 & 0.044 \\
HPWD1 & 100 & 0.959 & 0.951 & 0.938 & 0.982 & 0.948 & 0.915 & 0.958 & 0.820 & 0.970 & 0.964 & 0.028 \\
HPWD5 & 100 & 0.947 & 0.942 & 0.934 & 0.963 & 0.928 & 0.919 & 0.960 & 0.793 & 0.957 & 0.961 & 0.067 \\
HPWD10 & 100 & 0.948 & 0.948 & 0.944 & 0.967 & 0.917 & 0.915 & 0.948 & 0.793 & 0.963 & 0.937 & 0.044 \\
\hline
GRTS & 300 & 0.935 & 0.923 & 0.902 & 0.942 & 0.902 & 0.875 & 0.934 & 0.784 & 0.921 & 0.927 & 0.296 \\
LPM & 300 & 0.908 & 0.885 & 0.869 & 0.907 & 0.873 & 0.849 & 0.889 & 0.737 & 0.890 & 0.891 & 0.103 \\
SCPS & 300 & 0.922 & 0.873 & 0.865 & 0.899 & 0.868 & 0.848 & 0.885 & 0.742 & 0.901 & 0.894 & 0.070 \\
PWD1 & 300 & 0.928 & 0.921 & 0.898 & 0.937 & 0.898 & 0.900 & 0.918 & 0.780 & 0.926 & 0.930 & 0.055 \\
PWD5 & 300 & 0.900 & 0.888 & 0.852 & 0.903 & 0.859 & 0.850 & 0.880 & 0.733 & 0.891 & 0.877 & 0.075 \\
PWD10 & 300 & 0.892 & 0.876 & 0.859 & 0.895 & 0.856 & 0.825 & 0.883 & 0.720 & 0.877 & 0.886 & 0.045 \\
HPWD1 & 300 & 0.945 & 0.911 & 0.913 & 0.952 & 0.903 & 0.882 & 0.915 & 0.776 & 0.911 & 0.935 & 0.036 \\
HPWD5 & 300 & 0.904 & 0.902 & 0.887 & 0.903 & 0.870 & 0.840 & 0.878 & 0.723 & 0.889 & 0.883 & 0.066 \\
HPWD10 & 300 & 0.895 & 0.887 & 0.846 & 0.897 & 0.862 & 0.830 & 0.878 & 0.722 & 0.887 & 0.881 & 0.043 \\
\hline
GRTS & 600 & 0.903 & 0.869 & 0.852 & 0.900 & 0.860 & 0.836 & 0.872 & 0.730 & 0.898 & 0.888 & 0.298 \\
LPM & 600 & 0.853 & 0.840 & 0.806 & 0.838 & 0.823 & 0.780 & 0.817 & 0.679 & 0.845 & 0.837 & 0.109 \\
SCPS & 600 & 0.854 & 0.822 & 0.805 & 0.822 & 0.814 & 0.770 & 0.815 & 0.681 & 0.839 & 0.825 & 0.078 \\
PWD1 & 600 & 0.906 & 0.882 & 0.862 & 0.904 & 0.867 & 0.846 & 0.878 & 0.734 & 0.905 & 0.898 & 0.066 \\
PWD5 & 600 & 0.846 & 0.836 & 0.792 & 0.820 & 0.810 & 0.771 & 0.811 & 0.673 & 0.838 & 0.819 & 0.085 \\
PWD10 & 600 & 0.837 & 0.823 & 0.793 & 0.819 & 0.807 & 0.760 & 0.796 & 0.659 & 0.834 & 0.820 & 0.056 \\
HPWD1 & 600 & 0.892 & 0.862 & 0.858 & 0.899 & 0.868 & 0.840 & 0.884 & 0.734 & 0.893 & 0.889 & 0.052 \\
HPWD5 & 600 & 0.842 & 0.834 & 0.796 & 0.810 & 0.800 & 0.758 & 0.807 & 0.663 & 0.823 & 0.817 & 0.076 \\
HPWD10 & 600 & 0.815 & 0.806 & 0.775 & 0.781 & 0.791 & 0.741 & 0.795 & 0.638 & 0.806 & 0.798 & 0.053 \\
\hline
\multicolumn{2}{l|}{Area (\it{hec.})}  & \multicolumn{1}{c}{1688} & \multicolumn{1}{c}{2380} & \multicolumn{1}{c}{1764} & \multicolumn{1}{c}{692} & \multicolumn{1}{c}{3120} & \multicolumn{1}{c}{1452} & \multicolumn{1}{c}{1368} & \multicolumn{1}{c}{5976} & \multicolumn{1}{c}{2368} & \multicolumn{1}{c|}{1300} & \\
 \cline {1-12}
\end{tabular}
\end{center}
\end{table*}
In this Section the performance of the HPWD is empirically compared with respect to alternative {spatially balanced} designs via 10,000 sample replications, which have been carried out by using the free software environment for statistical computings R \cite{b49}. In particular, we used the following R packages: \texttt{BalancedSampling} \cite{b30}, and \texttt{spsurvey} \cite{b41}. For further details on the R codes that can be employed to randomly select {\it spatially balanced} samples, see \cite[ch. 7]{b06}.\\
As possible alternatives to the HPWD, we considered the GRTS \cite{b55}, the SCPS \cite{b29}, the LPM \cite{b31} and the PWD \cite{b05}. In addition, for PWD and HPWD, three values for the parameter $\gamma \in \{1,5,10\}$ were tried, indicated as suffix after the acronym of the design. To be comparable each other, all these alternative designs have been used setting the first order inclusion probabilities constant and equal to $n/N$.\\
The comparison between different designs has been performed by using the (RMSE) of the HT estimates as relative to the RMSE obtained when using a SRS design that is used, thus, as a scale factor to remove the known effects of the size of the population $N$ and of the sample $n$ on the sampling errors. It is worth noticing that in every simulation performed, as the HT estimator is unbiased, the RMSEs were always very close to the standard error of each design as the bias can be considered negligible.\\
The experiments carried out concern both real and artificial populations. The latter are two frames of size $N$=1,000 generated through point processes with two different levels of clustering of the units to control the distribution of the coordinates and with different spatial features of the response variable $Y\bigl(i\bigr)$. The coordinates $\{\vec{x}_1,\vec{x}_2\}$ are generated in the unit square $[0,1]^2$ according to a Neyman-Scott process with intensity of the cluster centers equal to 10 \cite{b63} with 100 expected units per cluster. Two different scale parameters for cluster kernel were used $\left \{0.005, 0.03\right \}$, representing respectively an highly clustered and a sparse population of spatial units (see Figure \ref{f:popsim}). For each frame four possible outcomes $Y\bigl(i\bigr)$ have been generated according to a Gaussian Markov Random Field with or without a spatial linear trend $\vec{x}_1+\vec{x}_2+\eta$, that explain approximately the $80\%$ of the variance of $Y\bigl(i\bigr)$. Spatial dependence of the errors $\eta$ is modeled by an exponential variogram with two intensities of the parameter $\left \{0.01, 0.1\right \}$, representing respectively medium and high dependence between units. To avoid the possible effects due to different variability, each population was finally standardized to the same mean $\mu_Y=5$ and standard deviation $\sigma_Y=1$.\\
To verify if $n$ has any effect on the efficiency, from each population, samples of different size $n\in \left \{10, 50, 200\right \}$ have been selected.
One of the two real populations used is the well known case study, largely debated and analyzed in the field of geostatistics, the Meuse River data--set available in the package \texttt{gstat}. It is a set of 155 samples of top soil heavy metal concentrations (ppm) used in our experiments as a population, along with a number of soil and landscape variables. The samples were collected in a flood plain of the river Meuse, near the {\it Stein} village (The Netherlands). Heavy metal concentrations are bulk sampled from an area of approximately $15m \times 15m$. In addition to the topographical map coordinates the  top soil concentrations of four metals have been used: cadmium, copper, lead and zinc. On these variables there is an high spatial autocorrelation as it is apparent from the variograms (Fig. \ref{f:meuse}). In this population the experiment consists of selecting samples of size $\{10,50\}$.\\
Land is very important for most biological and human activities, it is the main economic resource for agriculture, forestry, industries, and transport. The information collected on the land deals mainly with two interconnected concepts: land cover that refers to the biophysical coverage of land (e.g. crops, grass, broad-leaved forest, or build-up area) and land use that specifies the socio-economic use of land (e.g. agriculture, forestry, recreation or residential use). It is for this reason that land use and land cover data collected by the EU survey LUCAS has been chosen as the second real population.\\ 
The Land Use/Cover Area frame Survey (LUCAS) \cite{b27} is a task by EUROSTAT that was initially designed to deliver, on a yearly basis, European crop estimates and then became a more general agro-environmental survey.\\
Our population has been built considering 2012 as reference year and the region Emilia--Romagna in Italy as the area under investigation (see Figure \ref{f:lucas}). The regular grid $2km \times 2km$ of points constitutes our frame population. In the LUCAS experiment three different sample sizes have been adopted $\{100,300,600\}$ on data aggregated in 10 land cover classes.\\
The first check that must necessarily be done before performing the estimates is related to the assurance of having respected the $\pi_i$s, otherwise the HT estimator could provide extremely biased results. From the $CV\left ( \hat{\pi}_i \right )$s (\ref{eq:cvpi}) reported in Table (\ref{t:tabpi}) it is clear that the standardization of the distance matrix implied again as a result that the $\pi_s$s are approximately constant even if, with the exception of the LUCAS frame that lies on a regular grid, the error introduced is quite higher than that observed in regular grid populations. The spatial distribution of the population is thus an essential aspect to entail an efficient standardization of $\mathbf{D}_U$. In addition to the effects of $\gamma$ and $n$ already observed in regular grids it is possible to notice that a combination of a small sample size, an high tuning parameter and the presence of clustering in the coordinates of units in $U$ may worryingly reduce the accuracy of the $\pi_i$s. In such cases it would be advisable to proceed to a partition of the frame population in zones with a more homogeneous spatial distribution and then standardize the $\mathbf{D}_U$ independently within each stratum. It should also be considered that the $CV\left ( \hat{\pi}_i \right )$s that may seem high actually did not result in significant biases in the HT estimates proving its robustness to even medium-high deviations in the $\pi_i$s or in their estimated counterpart $\hat{\pi}_i$s. It is indeed necessary to remember that in this case we used only 10,000 sample replications instead of the 100,000 used for regular grids, and probably they are not sufficient to achieve reliable estimates for the inclusion probabilities.\\
Looking at Tables (\ref{t:tabsim}), (\ref{t:tabmeuse}) and (\ref{t:tablucas}), it can be noticed that generally the HPWD gives encouraging results as it seems to handle any existing spatial data structure and effectively treat it to locate units in the study region. Clustering of the population, presence of a spatial trend and of spatial autocorrelation imply a clear direct effect on the reduction of estimation variance even though their joint impact is obviously extremely moderate.\\
Especially a linear trend has been a valuable attribute to be exploited by our design even if very high variance decreases are also found in the presence of spatial autocorrelation, while the clustering of the population units slightly mitigate its capacity to spread the sample. Although it seems to be so sensitive to the occurrence of any of these properties it is also quite robust in case of their absence, as it always has a RMSE much smaller to that obtained using the SRS. Note that GRTS, SCPS and LPM have a similar behavior but with a lower gain in efficiency in all these situations, in particular GRTS that shows to achieve samples that are less {\it spatially balanced} than any other method.\\
The Meuse example confirms that HPWD is at least as efficient as PWD and that all the trends found in artificial populations are also verified on this case study. The results of the LUCAS data are also very encouraging. The gain in RMSE of the estimates compared with SRS is remarkable reaching and exceeding, when $n=600$ in some land cover classes, 25\%.\\
In any experiment the PWD is slightly but systematically more efficient than SCPS and LPM if used with $\gamma \geq 5$ and sometimes a bit less efficient when used with lower exponents. The HPWD proves to be an appreciable approximation of the PWD as it always has very similar RMSEs and, in some rare cases, surprisingly better. The increase of $n$ invariably entails an improvement in the performance of PWD and HPWD with respect not only to SRS but also to GRTS while the difference with SCPS and LPM seems to remain appreciably constant.\\
Table (\ref{t:tabctime}) reports the average CPU time in seconds of a 3,06 GHz Intel Core 2 Duo used by each of the algorithms to select a sample for different populations and sample sizes. For each method the time of both the developed version using R and C ++, when available (except for the GRTS, both versions are always available), is reported. For SCPS and LPM, the R code has been extracted from the available online material included in \cite{b29,b31}. The CPU time of LPM are reported for both the suggested methods: LPM1 is slower but more accurate (used to produce any results in this article) and LPM2 is faster but less accurate.\\
The extent to which the execution time matters, particularly if it highly depends on $N$ and $n$, is not secondary to the choice of the design. It is clear that among the examined procedures the GRTS is the more computationally intensive method and that the HPWD is sensibly quicker than the other distance based methods. The time spent to select a sample gradually increases with $n$ and only proportionally with $N$ mainly because the number of iterations is exactly $n$ while the time used by SCPS and the two versions of the LPM increases with $N^2$. To be honest, however, this comparison is strongly influenced by the computation of $\mathbf{D}_U$ that in PWD and HPWD is excluded from the computing times, since, having this matrix to be standardized in each application, its external calculation is required. In SCPS and LPM the distances calculation is instead included in the sample selection algorithm.\\
Finally we can be confident that HPWD can be effectively applied without extensive difficulties even to spatial populations of large size and the only limit can be represented by the amount of memory needed to store the distance matrix.
\begin{table*}
\caption{Average computing time (in seconds) to select a sample based on 1,000 replications for each design and for different population and sample sizes.}
\label{t:tabctime}
\begin{center}
\setlength\tabcolsep{3 pt}
\begin{tabular}{c|c|rrrrrr|rrrrr}
 & & \multicolumn{6}{c|} {R} & \multicolumn{5}{c}{C++} \\ 
$N$ & $n$ & GRTS & SCPS & LPM1 & LPM2 & PWD & HPWD & SCPS & LPM1 & LPM2 & PWD & HPWD \\
\hline
1000 & 100 & 0.182 & 3.114 & 8.348 & 2.008 & 0.243 & 0.014 & 0.058 & 0.013 & 0.007 & 0.009 & 0.000 \\
1000 & 300 & 1.941 & 3.094 & 8.457 & 2.010 & 0.374 & 0.039 & 0.058 & 0.013 & 0.007 & 0.027 & 0.002 \\
1000 & 600 & 57.008 & 2.962 & 8.137 & 1.960 & 0.594 & 0.075 & 0.058 & 0.013 & 0.007 & 0.040 & 0.003 \\
\hline
1500 & 100 & 0.116 & 6.728 & 17.698 & 4.393 & 0.376 & 0.019 & 0.135 & 0.028 & 0.015 & 0.036 & 0.007 \\
1500 & 300 & 0.848 & 6.665 & 16.662 & 4.161 & 0.598 & 0.059 & 0.133 & 0.028 & 0.015 & 0.014 & 0.002 \\
1500 & 600 & 23.195 & 6.668 & 17.627 & 4.375 & 0.952 & 0.112 & 0.134 & 0.028 & 0.015 & 0.030 & 0.029 \\
\hline
2000 & 100 & 0.104 & 11.920 & 31.587 & 7.757 & 0.504 & 0.026 & 0.246 & 0.050 & 0.026 & 0.049 & 0.001 \\
2000 & 300 & 1.088 & 11.848 & 31.586 & 7.682 & 0.791 & 0.074 & 0.246 & 0.050 & 0.026 & 0.023 & 0.001 \\
2000 & 600 & 5.267 & 11.718 & 31.084 & 7.661 & 1.264 & 0.147 & 0.245 & 0.050 & 0.026 & 0.038 & 0.003 \\
\hline
2500 & 100 & 0.115 & 18.511 & 49.659 & 12.036 & 0.642 & 0.031 & 0.392 & 0.078 & 0.041 & 0.014 & 0.001 \\
2500 & 300 & 1.108 & 18.304 & 49.919 & 12.051 & 1.019 & 0.094 & 0.392 & 0.078 & 0.041 & 0.051 & 0.011 \\
2500 & 600 & 5.021 & 18.318 & 49.666 & 12.053 & 1.652 & 0.185 & 0.392 & 0.078 & 0.041 & 0.062 & 0.013 \\
\hline
3000 & 100 & 0.134 & 26.623 & 71.000 & 17.255 & 0.769 & 0.038 & 0.574 & 0.112 & 0.059 & 0.058 & 0.001 \\
3000 & 300 & 1.418 & 26.412 & 71.197 & 17.343 & 1.242 & 0.115 & 0.573 & 0.113 & 0.059 & 0.028 & 0.005 \\
3000 & 600 & 4.688 & 26.532 & 66.318 & 16.284 & 1.972 & 0.223 & 0.564 & 0.113 & 0.059 & 0.054 & 0.011 \\
\hline
\end{tabular}
\end{center}
\end{table*}
\section{Conclusions}
\label{s:conc}
The use of space as a criterion for selecting sample units from a population is a solution that promises significant developments in ecological, environmental, agricultural and forestry surveys. In addition to collecting data on phenomena that are impossible or particularly expensive to observe with an exhaustive direct observation of the frame population, this type of surveys allows also to update existing lists frames (e.g. agricultural holdings), to use auxiliary information available only on a geographical basis (remotely sensed data), to facilitate some aspects of quality controls and to better define some concepts and nomenclatures for data dissemination (small area estimation).\\
Recent advances in geocoding that have led to a better understanding of the position of population units have thus extended the interest in these applications by the Institutes responsible for the production of official statistics. Typical households and business surveys have been based on archives that, in addition to many administrative data, include spatial information on each single unit of the population. It also results that these institutes have to update the survey techniques adopted so far. Among these, random sample selections have seen both theoretical and practical developments in methods and algorithms, allowing us to conduct surveys efficiently on spatial populations. Their motivation is still unclear and their framing is still debated within two approaches, based on the model or on the design, formally very different but that lead to logical and intuitive choices that can be reconciled. The selection with probability proportional to the distance between sampling units seems to go in this direction with the use of the PWD method. One of the main limitations of this studies lies on the fact that only the specification of homogeneous and isotropic spatial processes can be considered by using the distance matrix as a summary of the population spatial features. The population size, however, could be a big obstacle to be circumvented as this method is based on a computationally intensive MCMC approximation of the distribution $P(S)$ representing the sample design. Simultaneously increasing the amount of surveys on which to use these methods and information details on the phenomena to be investigated, both in terms of spatial resolution and size, it has become essential to propose a much faster alternative to PWD, the HPWD that can actually randomly select $S$ in exactly $n$ steps.\\
The results of the empirical studies show that the proposed HPWD design can compete in precision with the best {\it spatially balanced} designs analyzed, while also allowing large reductions in the computational effort required. It represents an important alternative to such designs when the survey requires, as it is often the case in real situations, the use of a very large spatial population. The results obtained seem to confirm that the simplicity of a draw--by--draw scheme may also be sufficient to handle many spatial aspects of the frame population and of the collected variables.\\
Some issues remain open for future research fundamentally linked to the theoretical derivation of $\pi_i$s and $\pi_{ij}$s when the HPWD or PWD are used. Their knowledge would enable to improve the study the theoretical properties of these designs and perhaps to derive more flexible alternatives that can be better adapted to the treatment of spatial samples repeated over time or to the selection of samples well--spread, stratified and simultaneously balanced on some auxiliaries.

%\begin{acknowledgements}
%\end{acknowledgements}

\end{document}